\documentstyle[11pt,adassconf]{article}
\begin{document}

\paperID{P2.3.9}

\title{Processing of 24 Micron Image Data at the {\it Spitzer} Science Center}

\author{Frank J. Masci, Russ Laher, Fan Fang, John Fowler, Wen Lee,
        Susan Stolovy, Deborah Padgett and Mehrdad Moshir}

\affil{Spitzer Science Center, California Institute of Technology,
       Pasadena, CA 91125, Email: fmasci@ipac.caltech.edu}

\contact{Frank Masci}
\email{fmasci@ipac.caltech.edu}

\paindex{Masci, F. J.}
\aindex{Laher, R.}
\aindex{Fang, F.}
\aindex{Fowler, J.}
\aindex{Lee, W.}
\aindex{Stolovy, S.}
\aindex{Padgett, D.}
\aindex{Moshir, M.}

\authormark{Masci et al.}


\begin{abstract}
The $24\micron$ array on board the {\it Spitzer} Space Telescope
is one of three arrays in the Multi-band Imaging Photometer
for {\it Spitzer} (MIPS) instrument. It provides
$5\arcmin.3\times5\arcmin.3$ images at a scale of $\simeq2\arcsec.5$ per
pixel corresponding to sampling of the point spread function
which is slightly better than critical ($\simeq0.4\lambda/D$).
A scan-mirror allows dithering of images
on the array without the overhead of moving and stabilizing
the spacecraft. It also enables efficient mapping of large areas of sky
without significant compromise in sensitivity.
We present an overview of the pipeline flow and reduction steps
involved in the processing of image data acquired with the
$24\micron$ array. Residual instrumental signatures not yet
removed in automated processing and strategies for hands-on
mitigation thereof are also given. 
\end{abstract}

\vspace{-5mm}
\section{Introduction}
\vspace{-3mm}

Since the launch of {\it Spitzer} in August 2003,
observations with the MIPS-$24\micron$ array have enormously extended
our understanding of the infrared Universe. The array has attained
sensitivities $\simeq1.5$ times better
than pre-launch estimates (see Rieke {\it et al.}, 2004 for a review),
allowing imaging of star forming regions and
high redshift galaxies to sensitivities and spatial resolutions approaching
factors of $\simeq10^{3}$ and $\simeq10^{2}$ respectively
better than {\it IRAS}.

The MIPS-24$\micron$ array is a $128\times128$ pixel
Si:As Blocked Impurity Band (BIB) detector and operates in a broad spectral
band extending from 21 to about $27\micron$. Pixels are continuously
and non-destructively read out every $\simeq0.52$ sec over possible
integrations ranging from $\simeq3$ to 30 sec. 
Due to bandwidth restrictions, the individual
samples are not downlinked. Instead, a line is fitted to the ramp
samples for each pixel using an on-board linear regression algorithm. The data
frames are downlinked in units called Data Collection Events (DCEs)
and are packaged by the Multi-mission Image Processing
Laboratory (MIPL) into two-plane FITS cubes. The first plane contains
the fitted slopes for each pixel and the second, the difference between
the first two reads in the ramp, referred to as the first-difference.
The first-difference frame effectively increases the dynamic range
for slopes derived from ramps which saturate.
To further limit data volume,
only first-difference values which exceed a nominal threshold
(set by the saturation level) are retained and downlinked.

The 24$\micron$ array has two data-taking modes. That just described is called
``SUR'' (for Sample-Up-the-Ramp) and is the primary science
mode for this array. The second mode is called ``RAW'',
where all ramp samples are downlinked and received as multi-plane FITS
cubes. RAW-mode is only used for engineering purposes. All raw pixel data
are represented in signed 16-bit integer format, with SUR-mode data in
units of Data Number/read-time (DN/read) and RAW-mode in DN.
The popular observing mode is scan-map\footnote{For a description of all
modes, see http://ssc.spitzer.caltech.edu/documents/SOM/} mode, and
results in the highest data volume with typically 11,000 DCEs
downlinked per day for the 24$\micron$ array alone.

To process such large data volumes, an infrastructure
of automated pipelines, running on a cluster of 34 CPUs has been
set up at Caltech's {\it Spitzer} Science Center (SSC) (e.g., Moshir, 2001).
This paper reviews the reduction steps used to process specifically
MIPS-$24\micron$ SUR-mode science data at the SSC. A summary of instrumental
residuals not yet corrected in automated processing, and which may appear
in products distributed to observers, is also given.

\vspace{-3mm}
\section{Pipeline Infrastructure Summary}
\vspace{-3mm}

Raw images undergo several stages of automated processing at the SSC
to produce the best calibrated products. The SSC is responsible
for archiving, distributing
data to observers, and maintaining current information on
instrument calibration. For the MIPS-$24\micron$ array, eight separate
pipelines have been developed for the removal of instrumental artifacts at
the DCE level, each specific to the particular data collection
mode or flavor of calibration product needed. These are as follows:

\begin{enumerate}
\item SUR-mode {\it \underline{science}} (input: two-plane FITS DCE of slope and difference).
\item RAW-mode {\it \underline{science}} (input: multi-plane FITS DCE of sample reads).
\item SUR-mode dark-current {\it \underline{calibration}} (input: ensemble of DCEs).
\item RAW-mode dark-current {\it \underline{calibration}} (input: ensemble of DCEs).
\item Electronic non-linearity {\it \underline{calibration}} (input: ensemble of RAW-mode DCEs).
\item Flatfield (non-uniformity) {\it \underline{calibration}} (input: ensemble of SUR-mode DCEs).
\item Latent-image flagging (input: preprocessed ensemble of BCDs).
\item Pointing reconstruction and Final Product Generation (FPG) on BCD.
\end{enumerate}

On ingestion of data, a pipeline executive ensures that calibration data
are processed first, before being employed in the reduction of regular science DCEs.
Calibration products are created from ensembles of input DCEs,
and a procedural database query tool using a set of predefined rules
has been developed (Laher \& Rector, 2005).
Calibration products are transferred to the science pipelines
using database queries handled by a software module called
CALTRANS (Lee {\it et al.}, 2005). The main product resulting from a DCE is a Basic
Calibrated Data product, or BCD, with raw pointing and distortion information
attached to its FITS header. The final (Post-BCD) processing steps include pointing
refinement using astrometric matching (Masci {\it et al.}, 2004), and mosaicking
of ensembles of BCDs to provide seamless final products (Makovoz {\it et al.}, 2005).
Post-BCD processing is not discussed in this paper.

For each processed $24\micron$ science DCE, ten associated BCD image products are
archived at SSC, a subset of which are distributed to users. 
The archived set includes uncertainty images, bit-mask images
which summarize the processing status for each pixel (both for the slope and first-difference
planes), and processing log files. For a full description, see the
{\it MIPS Data Handbook}\footnote{http://ssc.spitzer.caltech.edu/mips/dh/}. 
The above suite of pipelines has also been implemented into
offline stand-alone versions to facilitate testing and validation of algorithms
before deployment in operations.

\vspace{-3mm}
\section{BCD Processing Summary}
\vspace{-3mm}

\begin{figure}
\vspace{-0.5cm}
\epsscale{1.0}
\plotone{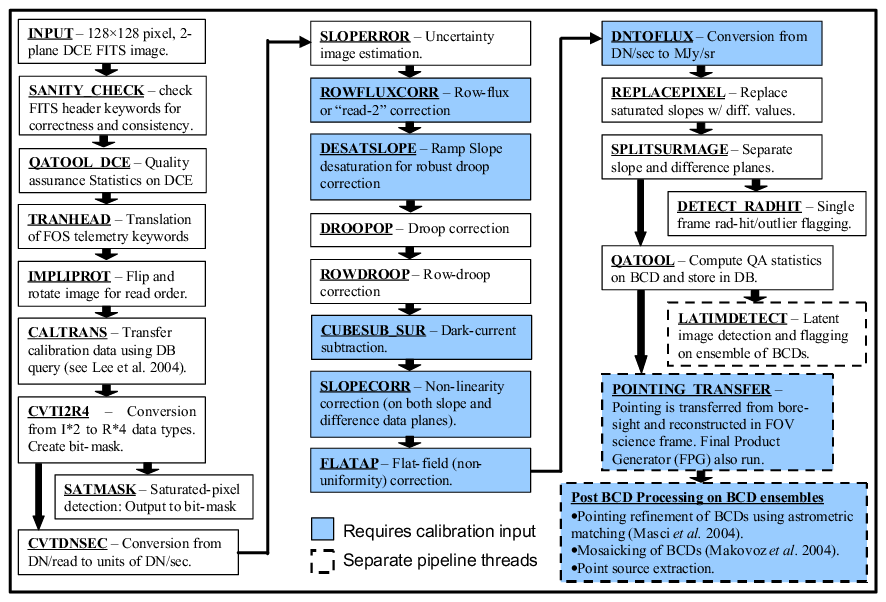}
\vspace{-0.8cm}
\caption{Processing flow in MIPS-$24\micron$
primary science pipeline.\label{fig1}}
\end{figure}

Figure~\ref{fig1} gives the ordering of reduction steps in the primary
$24\micron$ science pipeline. Processing algorithms were designed
in collaboration with the MIPS Instrument Team. For a review, see
Gordon {\it et al.} (2004). Here we give an overview of
some of the more important reduction steps and
instrumental signatures unique to the $24\micron$ array.

The first step which modifies pixel values is CVTI2R4. This converts
the signed 16-bit native raw image data into 32-bit floating point.
This step also applies a truncation correction to the on-board computed
slopes by adding a constant of 0.5 DN/read to every pixel
in the slope plane.

The next step is detection and flagging of ``soft''
and ``hard'' saturated pixels (SATMASK). Soft saturation is where the samples
saturate somewhere along the ramp, therefore biasing slope estimates.
By saturation, we mean that the samples become pegged to the maximum
value ($+32768$) as allowed by the on-board Analog-to-Digital Converter (ADC),
which corresponds to $\simeq30\%$ full well.
Soft saturated pixels are detected by simply thresholding pixels in the
first-difference plane which are above a nominal value. For a 10 sec exposure,
this value is $\simeq3000$ DN/read. Hard saturation is when
the first sample, and all samples thereafter become pegged to the maximum
ADC value. This is detected if both the derived on-board slope and
first-difference are zero. Both soft and hard saturated pixels are flagged
in a processing status mask and propagated downstream.

The SLOPERROR step initializes
an uncertainty image using a robust noise model as applicable
to slope data. This includes read noise, Poisson
noise, and an estimate of the correlation between samples along
a ramp from which the slope is derived. For a general overview
of uncertainty propagation in SSC pipelines, see Moshir {\it et al.}, (2003).

The ROWFLUXCORR step corrects a signature known as
the ``read-2'' effect. This effect describes a bias
introduced into the on-board slope measurement from a small additive offset
in the second sample of every ramp. This offset is seen to vary
across the array at the 0.2\% level, primarily in the cross-readout (row) direction.

The DESATSLOPE is used to de-saturate slope pixels that were
flagged for saturation from above (SATMASK). The de-saturated
slope pixels are not propagated downstream. They are only used
for robust computation of an effect known as ``droop''.
Droop is an extraneous signal that is added to each pixel (at the $\sim10\%$ level)
by the readouts. It is computed and removed by the DROOPOP module.
Droop is directly proportional to the total number of counts on the array,
including counts which are present above the ADC saturation level.
For saturated pixels, slopes are underestimated, and thus for the
purpose of computing droop, it is necessary
to estimate slopes that would result from ramps which would continue
beyond the saturation point. Slopes are de-saturated using the
first-difference value and the non-linearity model.
The ROWDROOP module corrects a second-order effect which similar to droop
but whose signal in a pixel depends on the total signal from all pixels
in its row.

The three standard calibrations are next. The dark current on each
pixel is removed by subtracting a dark calibration
image from both the slope and first-difference planes (CUBESUB\_SUR).
The dark calibration is computed by performing a symmetric outlier-trimmed average
of a few hundred DCEs taken with the scan-mirror in the dark position.
The dark current is small, and values range within 0-3 DN/sec.
Correction for electronic
non-linearity is performed by the SLOPECORR module.
The non-linearity is accurately described by a quadratic up to ADC
saturation, and the deviation from linearity
is typically 10-15\%. Flat-fielding is next (FLATAP).
Flatfield calibrations exhibit maximum deviations of $\simeq20\%$ from flatness,
and are primarily due to dark spots and low-level ``blotchiness''
in DCEs from absorption by debris on a pick-off mirror. 
The position and shape of these debris artifacts
depend on the angle and scan-rate of the scan-mirror respectively.
To correct for these, scan-mirror-dependent flatfields
are created offline following each initial campaign
processing run and deployed on the operations system before
reprocessing of science data.

The final steps include flux-calibration (DNTOFLUX)
and pixel replacement (REPLACEPIXEL). This last ``pixel-modifying'' step
looks for all pixels that were flagged for saturation in the mask
(from SATMASK), and replaces slope values with
first-difference values in the {\it primary BCD product}. The original processed
slope image (with no pixel replacement) is retained as an ancillary product.

\vspace{-3mm}
\section{Possible Instrumental Residuals}
\vspace{-3mm}

Broadly speaking, there are three instrumental signatures that may
remain in $24\micron$ BCD products after automated processing with the
S11.0 version of SSC pipelines. These can
be ameliorated with further hands-on processing, although
we expect to automate the corrections once
they are sufficiently characterized.

The first residual may arise from inaccurate,
scan-mirror dependent flat-fielding. As described above, the
process is not yet fully automated. Mismatches between dark spot
(debris-artifact) positions and corresponding actual mirror angles can occur.
These mismatches lead to bright and dark residual patterns
in BCDs, and given sufficient data, can be removed by re-creating
flatfields from the BCDs and performing a self-calibration.
For the most part, the scan-mirror-dependent flat-fielding is giving
excellent results, although observers are encouraged to report
anomalous cases to the SSC.

The last two residuals occur when bright, saturating
sources are observed. The first is called ``readout saturation'' and
occurs when a saturating
cosmic ray or source depresses the output of a single readout channel.
Since there are four readout channels, this gives the appearance
of a ``jail-bar'' pattern. This has been characterized as a
multiplicative effect and can be corrected by scaling the
affected readout columns with a median of the other three unaffected readouts.
This is only possible of course if
the background doesn't show complex structure. The second
effect is when the saturating source is bright enough to leave
``dark'' latents in many subsequent images.
They appear dark because the slopes are fitted to
saturated ramps, which turn out to be lower than average
on the array. Given a sufficient number of frames with no complex
structure, dark latents can be corrected by creating time-ordered
sequences of self-calibration flats and dividing these into the BCDs.

Overall, the behavior of the $24\micron$ array since launch
can be described as excellent. Unless bright saturating sources
are inadvertently observed, very few instrumental residuals are present.
Future work will focus on reducing the residuals just described. 

\acknowledgments
This work was carried out at the {\it Spitzer} Science Center, with 
funding from NASA under contract 1407 to the California Institute of
Technology and the Jet Propulsion Laboratory.


\begin{references}
\vspace{-3mm}
\reference Gordon, K. D., Rieke, G. H., Engelbracht, C. W., et al.\ 2004, \pasp, (in press)
\reference Laher, R., Rector, J.\ 2005, \adassxiv, \paperref{FW.1}
\reference Lee, W., Laher, R., Fowler, J., Masci, F., Moshir, M.\ 2004, \adassxiv, \paperref{P2.2.5}
\reference Makovoz, D., Khan, I.\ 2005, \adassxiv, \paperref{O10.3}
\reference Masci, F. J., Makovoz, D., Moshir, M.\ 2004, \pasp, 116, 842
\reference Moshir, M.\ 2001, \adassxi, 336
\reference Moshir, M., et al.\ 2003, \adassxii, 181
\reference Rieke, G. H., Young, E. T., Engelbracht, C. W., et al.\ 2004, \apjs, 154, 25
\end{references}
\end{document}